\documentclass{article}

\usepackage{PRIMEarxiv}

\usepackage[utf8]{inputenc} 
\usepackage[T1]{fontenc}    
\usepackage{enumerate}
\usepackage{graphicx}
\usepackage{amssymb}
\usepackage{longtable}
\usepackage{multirow}
\usepackage{appendix}
\usepackage[T1]{fontenc}
\usepackage{lscape}
\usepackage{lineno}
\usepackage{amsmath}
\usepackage{caption} 
\usepackage{subcaption} 
\usepackage{changes} 

\usepackage{url}
\usepackage{hyperref}
\usepackage[none]{hyphenat}

\title{IoT Anomaly Detection Methods and Applications: A Survey
\thanks{\textit{\underline{Citation}}: 
\textbf{Ayan Chatterjee and Bestoun S. Ahmed. \textit{Iot anomaly detection methods and applications: A survey}.
Internet of Things, 19:100568, 2022. DOI:10.1016/j.iot.2022.100568}} 
}

\author{
  Ayan Chatterjee \\
  Dept of Mathematics and Computer Science, Karlstad University, 651 88 Karlstad, Sweden \\
  \texttt{ayan.chatterjee@kau.se} \\
   \And
  Bestoun S. Ahmed \\
  Dept of Mathematics and Computer Science, Karlstad University, 651 88 Karlstad, Sweden\\
  Dept of Computer Science, FEE, Czech Technical University in Prague, Czechia \\
  \texttt{bestoun@kau.se} \\
}

\begin{document}
\maketitle

\begin{abstract}
Ongoing research on anomaly detection for the Internet of Things (IoT) is a rapidly expanding field. This growth necessitates an examination of application trends and current gaps. The vast majority of those publications are in areas such as network and infrastructure security, sensor monitoring, smart home, and smart city applications and are extending into even more sectors. Recent advancements in the field have increased the necessity to study the many IoT anomaly detection applications. This paper begins with a summary of the detection methods and applications, accompanied by a discussion of the categorization of IoT anomaly detection algorithms. We then discuss the current publications to identify distinct application domains, examining papers chosen based on our search criteria. The survey considers 64 papers among recent publications published between January 2019 and July 2021. In recent publications, we observed a shortage of IoT anomaly detection methodologies, for example, when dealing with the integration of systems with various sensors, data and concept drifts, and data augmentation where there is a shortage of Ground Truth data. Finally, we discuss the present such challenges and offer new perspectives where further research is required.
\end{abstract}


\section{Introduction} \label{sec:introduction}

The Internet of Things (IoT) enables sensors\footnote{Throughout this paper, the term `sensors' refers to sensors or devices attached to an IoT environment, which includes sensors that transmit data to edge, fog, or cloud architectures in an MLOps setting.} and smart objects to communicate without the direct involvement of human agents, necessitating near real-time processing \cite{IoTIntro2019}. Any data analytics performed via IoT requires the development of novel methodologies to work within the limited computational budget. One type of data analysis that looks for unusual states within the system is anomaly detection, also known as outlier detection or event detection. The anomaly detection algorithms are checkpoints for the incoming traffic at various stages, ranging from the IoT network level to the data center. In the latter case, there is a high demand for reliable detection for data cleaning \cite{ReviewOutlier2021} and classification purposes \cite{IoTSmartHomeArchi2021}.

The significance of anomaly detection is that anomalies in IoT data, which occur sparsely, can yield crucial actionable information in various sectors, including medicine, manufacturing, finance, traffic management, and energy. Anomaly detection in IoT, for example, is employed in the betting and gambling sector to detect insider trading by analyzing trade activity patterns \cite{horseBetting2021}. On the other hand, industrial machines use a detection algorithm to ensure production safety \cite{machineMonitor2020}.

Currently, most anomaly detection methods in the IoT involve significant human engagement and optimizations for local solutions. In theory, an anomaly is simple to comprehend, and a domain expert will spot anomalous data if given enough time. However, there are several difficulties in developing an automated model in an IoT environment. It is challenging and not always possible to define and categorize all types of anomalous data correctly, especially when labeled training data are only partially available/not available. In many fields, the notion of normal behavior is constantly changing and evolving. One such example is a change in household occupancy, which results in a change in electricity demand \cite{household2021}. Furthermore, data often contain noise, and when the signal-to-noise ratio is low, the magnitude of noise resembles true anomalies. The complexity increases as the number of interconnected systems grows and the variety of input data types.

IoT anomaly detection has a wide range of applications outlined in this paper, some of which are more developed than others (such as network security), and others that have the potential for growth. The literature on the subject of anomaly detection in the IoT is extensive and diverse.  However, the discipline is still in its early stages, and the number of articles published on this topic is expanding. Such frequent changes require updated literature. This survey adds to the earlier research and provides a current picture of recent developments and different domains of the IoT anomaly detection application. Additionally, this paper expands on the categorization of anomaly detection algorithms and conducts an extensive keyword trend analysis to demonstrate the year-over-year literature concentration.

The structure of the paper is as follows: Section \ref{sec:categorisation} provides a discussion on the background of anomaly detection algorithm in IoT and related work. The subsequent sections \ref{sec:search} and \ref{sec:recentAdvancements} present the search strategy for this paper and look at recent developments not described in previous reviews to the best of our knowledge. The final section \ref{sec:conclusion} includes new perspectives and concluding remarks.

\section{Background and related work}\label{sec:categorisation}

\subsection{Definition and categorization of IoT anomaly detection}

\begin{figure}[!ht]
    \centering
    \subfloat[]{\includegraphics[scale=0.32]{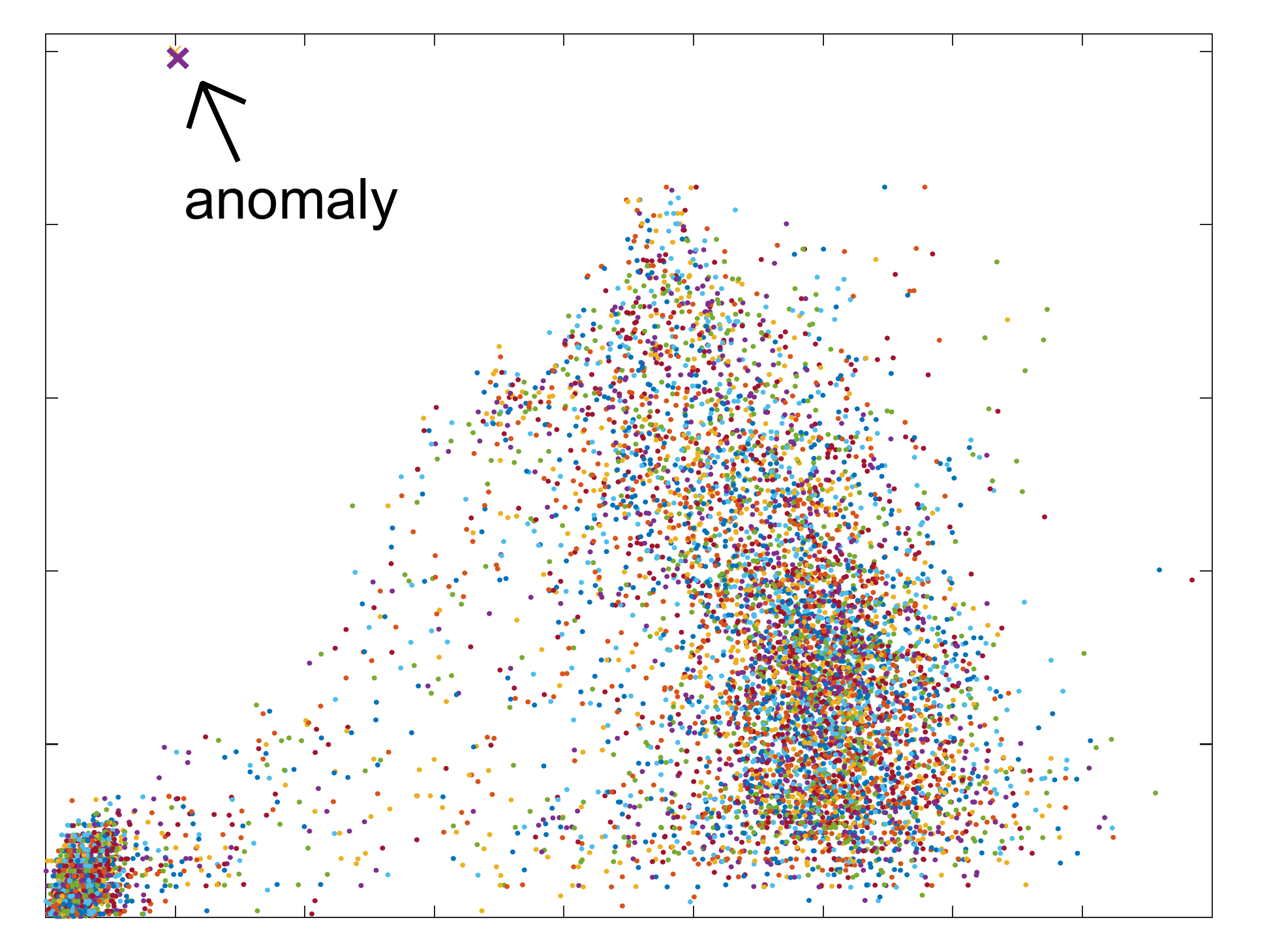}}
    \subfloat[]{\includegraphics[scale=0.32]{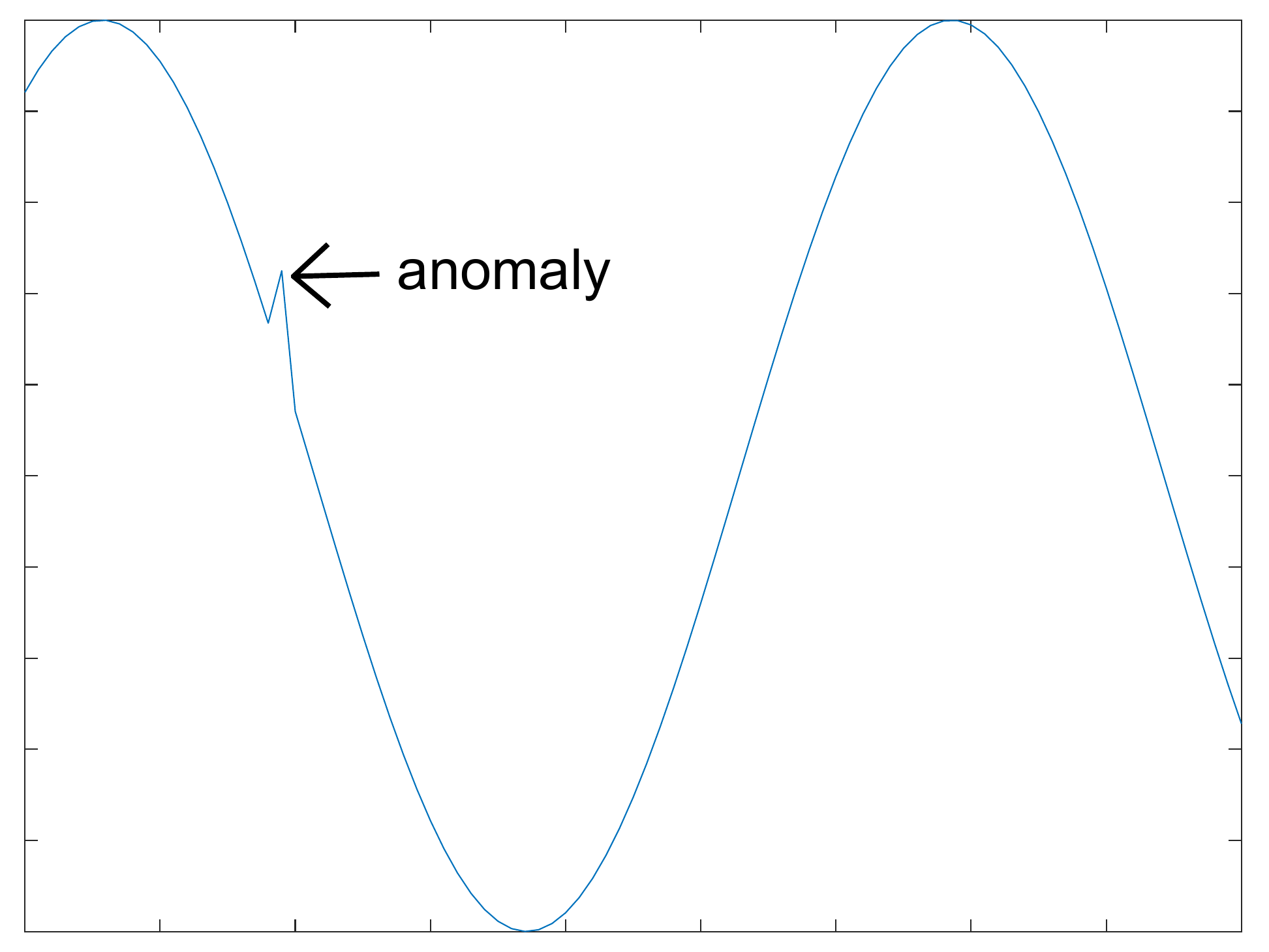}}
    \caption{A diagram exhibiting illustrated examples of anomaly.}
    \label{fig:anomalyEgs}
\end{figure}

An anomaly is a data point that is not associated with the predicted behavior in a modeled system\footnote{Anomalies in a simple, intuitive explanation are: Given a set of Lego (Lego is a trademark of The Lego Group) pieces, where the building instructions are withheld/unknown from the start, the task of a machine learning model is to generate a set of instructions and construct an object or a character. The remaining pieces that do not fit the trained model are anomalies.} Anomalies are rare events or observations that deviate significantly from conventional behavior or patterns observed in a single data point, a specific context or slice of time (for example, a season or quarter), or the entire dataset. In principle, anomalies result from external factors, such as sensor failure or external attack, and the purpose of a detection algorithm is to identify where an anomaly has occurred and classify/infer the cause. In the binary classification of an anomaly, the approximation model that best fits the expected data behavior is crucial. Furthermore, the complexities of many situations require a distinct detection strategy for each application \cite{IoTBook2020}. Example anomalies are shown in Figure \ref{fig:anomalyEgs}.

\begin{figure}[!ht]
    \centering
    \includegraphics[scale=0.48]{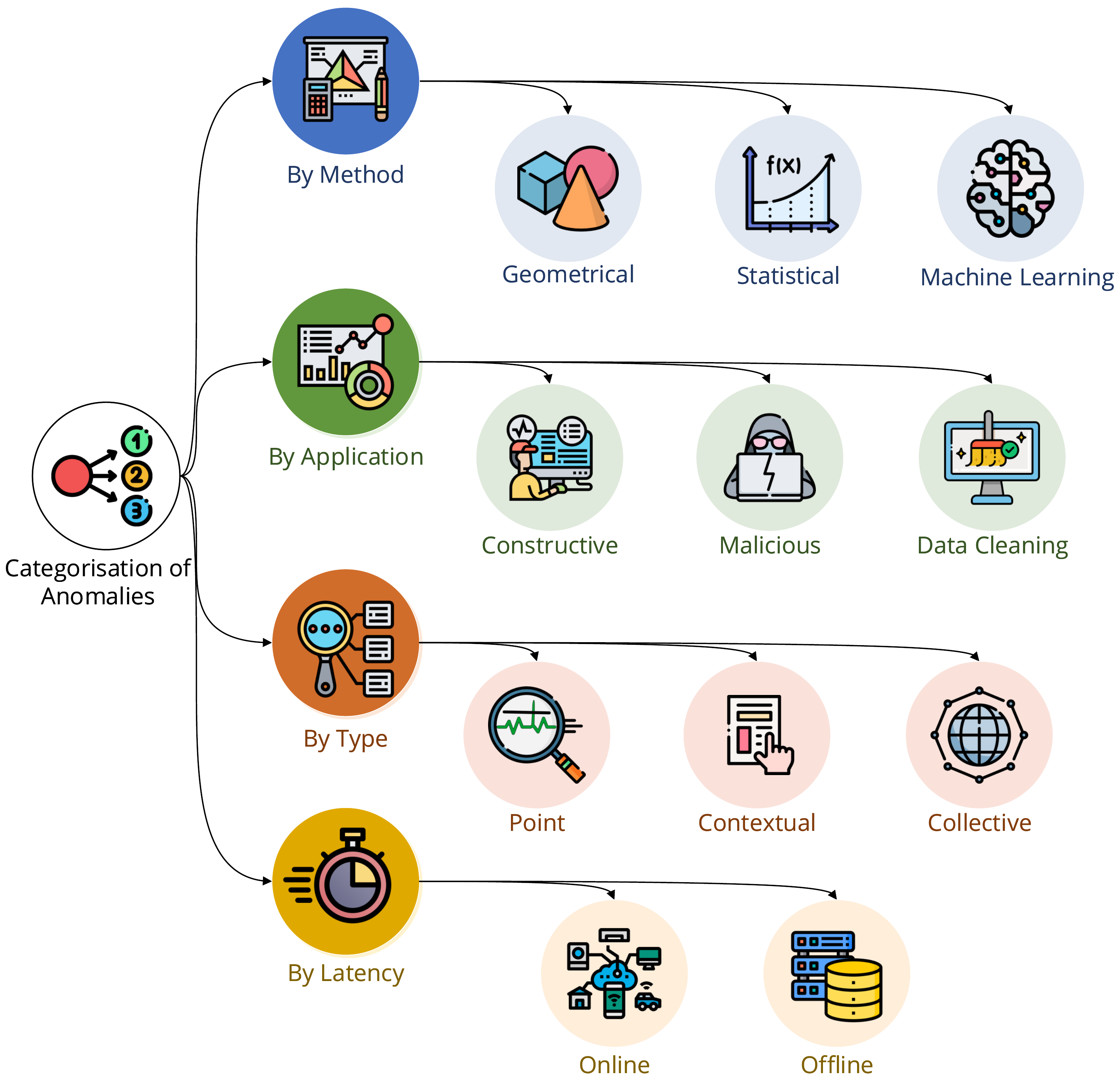}
    \caption{An info-graphic overview of the categorization of anomalies.}
    \label{fig:anomalyTypes}
\end{figure}

An IoT anomaly detection method is classified into four categories by combining the classifications from previously published articles such as Fahim and Sillitti \cite{review2019}, and Cook et al. \cite{review2019ts}. They are categorized based on how they approach the problem, how they are applied, the type of method, and the latency of the algorithm. An illustrative overview of the four categories is displayed in figure \ref{fig:anomalyTypes}. This section briefly describes the categorization of anomalies and some of the conventional approaches used in IoT.

\begin{enumerate}[i]
\item By Method: The approaches are geometrical, statistical, or machine learning. Geometrical methods are based on the assumption that when distance- and density-based strategies represent a given dataset, the expected and anomalous data are separated. In a set of data points, the idea behind isolation or density-based techniques is that anomalies appear in sparse regions. These methods use either a static or a dynamic threshold `t' on the estimated distance `d' to classify anomalies, which is given as:

\begin{align}
d = \left \{ \begin{array}{cccc}
    < t& ,& \text{Normal (under threshold)}\\
    > t& ,& \text{Anomaly (above threshold)}
\end{array}
\right.
\label{eq:thres}
\end{align}

Statistical methods such as the minimal volume in \cite{minVolume2016} try to model normal data using mathematical models and distributions. The minimal volume approach aims to create an n-dimensional simplex around the given data cloud (ground truth), where the objective function is to minimize the volume occupied while maximizing the ground truth data points. An anomaly is defined as any data that do not fit the simplex. Another is a forecasting approach called exponential smoothing \cite{es2018}. This technique anticipates the future data point using prior data points and a smoothing parameter. Anomalous data obtained from statistical methods are those that deviate from the established model. Conventional geometrical and statistical methods are backed by a substantial body of research and rely on a thorough understanding of the ground truth. These strategies do not reward many real-world instances in which data models are very time-dependent. Therefore, data-driven machine learning and deep learning solutions are required, which allow for more flexible modifications.

The third subcategory is machine learning and deep learning models, which have increased in publication frequency in recent years. The nature of the data supplied determines the selection of the model \cite{dlRorschach2018}. For example, long short-term memory (LSTM) and transformer models prefer sequential data inputs such as audio, video, and time series \cite{sequential2021}. On the other hand, the convolutional neural network (CNN) and autoencoder (AE) prefer non-sequential data types, such as image input \cite{healthcare2016}. The algorithms attempt to distinguish between normal and anomalous behavior by establishing a decision boundary, such as with the support vector machine (SVM) classifier \cite{oneClassSVM2021} or future forecasting values in streaming data \cite{streaming2018} with LSTM networks \cite{lstm2020}. Depending on the availability of training labels, these approaches are supervised, semi-supervised, self-supervised, or completely unsupervised \cite{MLmethods2019}.

\item By Application: The three routes of anomaly categorization by an application are constructive, destructive, and data cleaning. Constructive applications are productive or beneficial in nature and provide value to the world, such as monitoring the elderly's daily behavior for fall prevention in \cite{ElderlyFallDetection2017} using image descriptors and comparing the performance between multilayer perceptron (MLP), k-nearest neighbors (KNN), and SVM classifiers. Other examples include Lu et al.'s work \cite{uav2018} using reinforcement learning for various unmanned aerial vehicle (UAV) applications, including smart farming, and Nguyen et al. \cite{DIoT2019} using a federated learning approach for smart home applications. Destructive applications are intended to disrupt daily operations to accomplish questionable financial gain, an intention to harm the network and application data flowing through the IoT network, or disrupt critical business practices. These applications negatively impact society; for example, the paper by Alsheikh et al. \cite{SecurityTrend2021} surveys different IoT cyberattacks and the latest developments in IoT security. Such applications require research into solutions, such as RAPPER \cite{RAPPER2020} and NBaIoT \cite{NBaIoT2018} using an AE, for prevention or steps taken before the illicit incident and detection or actions taken after the incident. Finally, data cleaning or data cleansing applications, such as DeepAnT \cite{DeepAnT2019} using deep CNN, remove unwanted data spikes and sensor noise from the input signal.

\item By Anomaly Type: One of the most frequently encountered types is the circumstance-specific type, which is the point, contextual, and collective. An anomaly is a point anomaly if a single data point deviates from the expected behavior. An example is the detection of credit card fraud \cite{ccfraud2021}. A contextual anomaly is an instance that could be considered anomalous in some specific context. This means that comparing multiple perspectives on the same data point does not always result in anomalous behavior. A contextual anomaly is detected when both contextual and behavioral characteristics are considered together. For example, in the case of traffic offences, the anomalies vary depending on the geolocation information \cite{driving2020}. Unlike a point or contextual anomaly, the final type of anomaly, collective anomaly, looks at the entire dataset. An example of a collective anomaly is the use of electrocardiograms to monitor and detect anomalies or problems in the human heart \cite{ecg2020}.

\item By Latency: The latency and scalability of a detection algorithm determine whether the method is executed on the fly during the data collection stage or in a later stage of storage. An online algorithm can serially process information, with a single data-point or a window, without having access to the complete input. Conventional online geometrical and statistical approaches include the previously mentioned distance-based, density- and deviation-based, and angle-based techniques. Examples of online methods include the IoT-Keeper by Hafeez et al. \cite{IoTKeeper2020}, which uses fuzzy C-means, and Hedde et al. \cite{Ensemble2015} with an ensemble approach. On the other hand, offline algorithms have access to complete data. They involve relatively computationally expensive and sophisticated algorithms to solve the problem in a reasonable amount of time. However, it is essential to note that in recent years, algorithms such as the paper by Wu et al. \cite{LSTMBayes2020}, which uses LSTM and Gaussian Naive Bayes, and other models mentioned above, complete the model training process offline and deploy the model online.
\end{enumerate}

\subsection{Related work and our contribution}

A variety of articles on anomaly detection in IoT are published, ranging from expert insight on time-series methods by Cook et al. \cite{SurveyTimeSeries2020}, an introductory summary of the detection methods released between 2000 and 2018 by Fahim and Sillitti \cite{Survey2019}, an examination of sensor faults and outliers by Gaddam et al. \cite{SurveySensorOutliers2019}, and a comprehensive investigation of the work done in the early twenty-first century by Chandola et al. \cite{IndepthSurvey2009}. The most prevalent forms of surveys are data-specific, application-specific, or method-specific publications. For example, two recent surveys, a paper by Braei and Wagner \cite{SurveyUnivariate2020} and another by Mozaffari and Yilmaz \cite{SurveyOnlineMultivariate2019}, examined the use of univariate and multivariate data, respectively. Other examples include the survey by Santos et al. \cite{SurveySmartCityApps2018} that examines smart city applications on 5G wide area networks, a review paper on IoT security applications by Ahmad and Alsmadi \cite{SurveySecurity2021}, and anomaly detection using deep learning approaches in a survey by Chalapathy and Chawla \cite{SurveyDeepLearning2019}. Yassine et al. \cite{SurveyAIEnergy2021} provide a detailed examination of the methodologies, situations, and computing platforms used in the energy industry.

Despite the fact that this is not a systematic review paper, we did follow many of the techniques, but with some variations. For example, instead of using customized text searches, we developed a bubble chart to visually identify key research areas. Furthermore, based on our search, we found that the developments from 2019 and beyond are insufficient in other review papers. For this reason, our search criteria include papers published between January 2019 and July 2021. Our aim is to discover various IoT anomaly detection applications and report on recent findings. This paper complements the previous literature by:
\begin{enumerate}[i]
    \item examining publication trends, recent methods, and unique applications in IoT anomaly detection, and
    \item identifying present-day issues and challenges.
\end{enumerate}

\section{Search strategy} \label{sec:search}

\subsection{Extraction of keyword trends}
\begin{landscape}
\begin{figure}[!ht]
    \centering
    \includegraphics[scale=0.33]{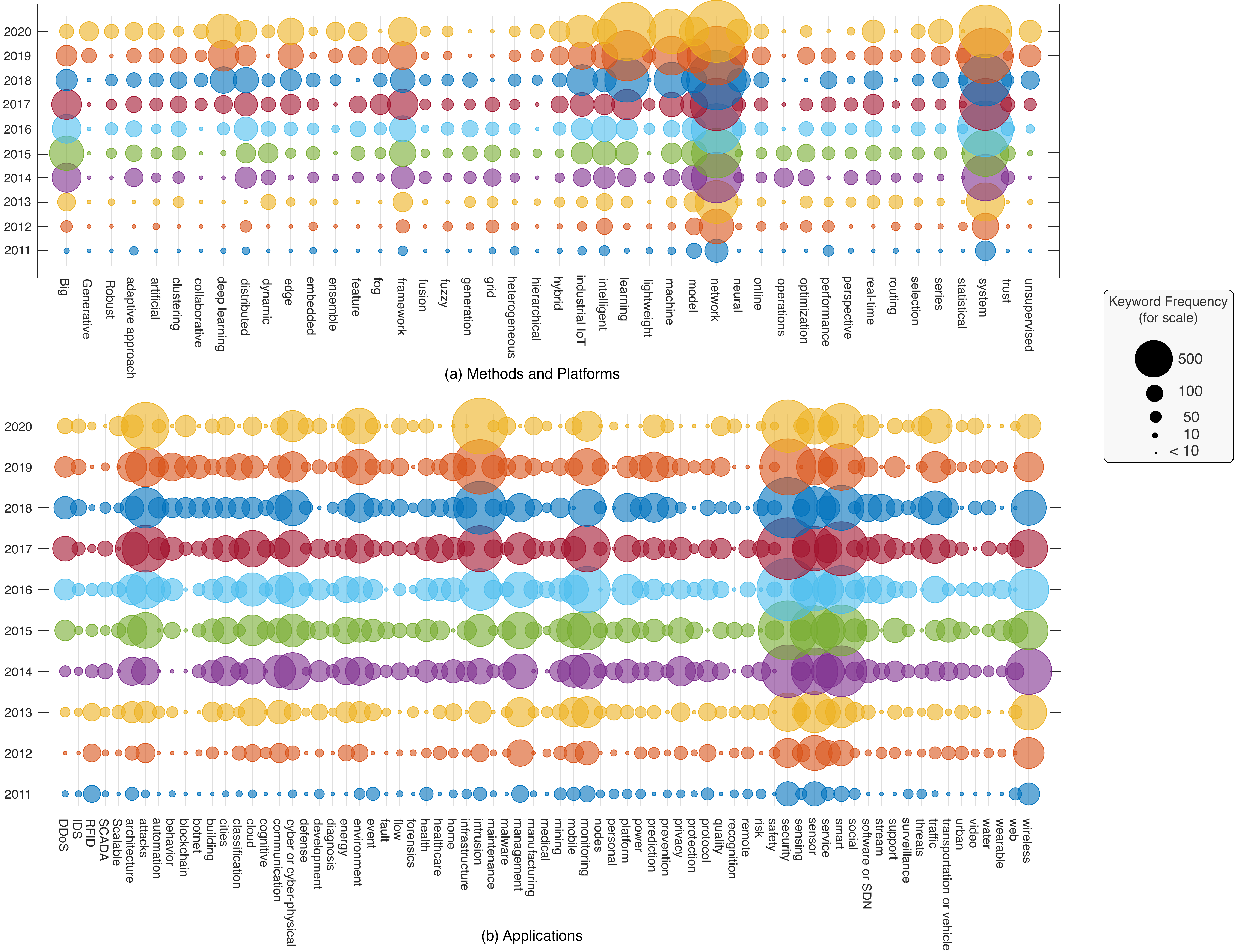}
    \caption{A chart displaying the number of times each keyword is used in article titles. This data is extracted from 8348 unique articles in IoT anomaly detection published between 2011 and 2020.}
    \label{fig:keywordsBubble}
\end{figure}
\end{landscape}

In this review, we include only papers published from January 2019 to July 2021. However, to better understand the trends in IoT anomaly detection research, we will extract search keywords from the titles of previous decade publications. We searched Google Scholar using the Publish or Perish tool from \cite{publishperishtool}. The search criteria were for all article titles in this domain with keywords `iot' and `anomaly' published between 2011 and 2020 and excluded duplicate titles. After merging the titles into a string, we extracted the unique keywords and their frequency of occurrence. Subsequently, generic terms such as `an', `the', and `novel' were eliminated from the text. We then sorted the remaining keywords into two categories: (a) methodologies and platforms and (b) applications. Figure \ref{fig:keywordsBubble} shows a bubble chart of publication keywords. Keywords with larger bubbles denote a higher frequency of publications, and smaller bubbles indicate a potential for an increase in contributions. From the figure, we discovered that the methods and platform keywords, which include `machine learning', `deep learning', `distributed', `edge', `framework', `industrial IoT', `intelligent', `network', and `system' all become increasingly relevant for the community. From the applications side, they are `architecture', `attacks', `cyber', `intrusion detection', `monitoring', `security', `sensor', `smart devices', and `wireless'.

\subsection{Papers' selection criteria}
\begin{figure}[!ht]
    \centering
    \includegraphics[scale=0.52]{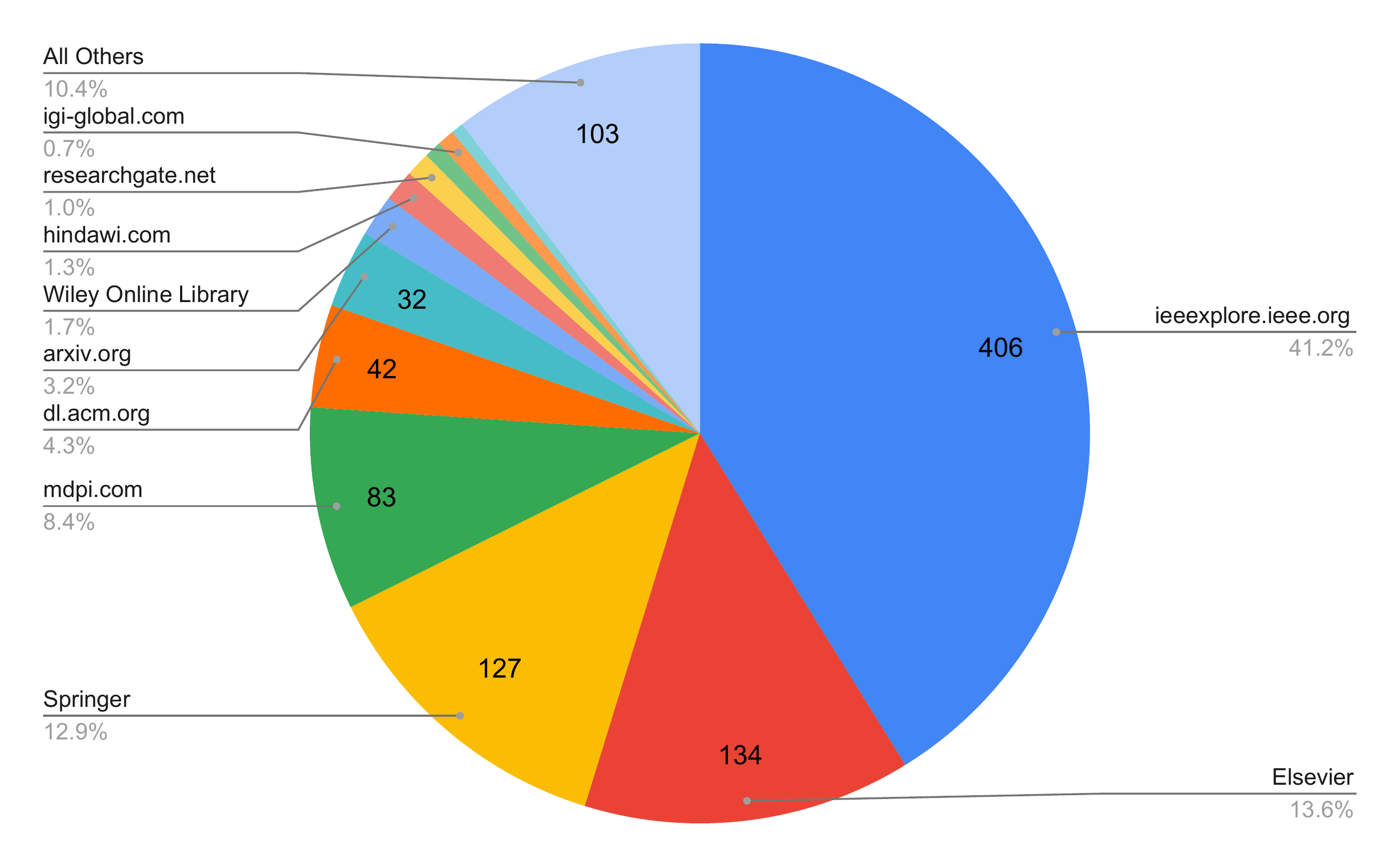}
    \caption{A pie chart illustrating the frequency of articles produced by publisher type between January 2019 and July 2021.}
    \label{fig:publisherPie}
\end{figure}

Following the extraction of the keyword trends, we performed a thorough search across established journals and conferences, in the order of the most recently published articles, using the search keywords `iot', `anomaly detection', and each of the trending keywords. There were searches in IEEE Xplore, Elsevier ScienceDirect, MDPI, Google Scholar, and arXiv. A total of 2242 articles were collected, which were published between January 2019 and July 2021. Duplicate papers and review publications were then removed using the exclusion keywords `survey', `literature review', `case study', `reviews', `systematic review', and `this review'. The titles of the remaining 1670 manuscripts were rapidly assessed, beginning with the relevance recommendations provided by the search engines mentioned and excluded patents and citation results. At this stage, there were 986 unique articles for further evaluation. Graphing the publishers of the remaining papers in a pie chart in figure \ref{fig:publisherPie} reveals that IEEE, Elsevier, Springer, and MDPI account for 76.1 percent of all recent publications.

The remaining articles were then filtered with the inclusion keywords `experiments', `evaluation,' and `results.' This allowed us to identify publications that had included experimental results as part of their
 work. Finally, the abstracts, discussions, and conclusions of the remaining 194 manuscripts were examined to identify papers relevant to the scope of this review. When determining the relevance of the content, the factors taken into consideration are as follows:
\begin{enumerate}
    \item This article made a notable contribution to the field of IoT anomaly detection. Although the search criteria specified inclusion and exclusion keywords, publications that were not directly related to this paper's context were excluded.
    \item The publication targeted a unique application domain. Multiple articles in the same application domain, such as botnet attack detection, were narrowed down to the best of our ability, selecting the most recent and higher Google Scholar search rank.
\end{enumerate}
Following the filtering process, we discuss the selected 64 articles published in the previous three years.

\section{Recent advancements} \label{sec:recentAdvancements}

\begin{figure}[!ht]
    \centering
    \includegraphics[scale=0.8]{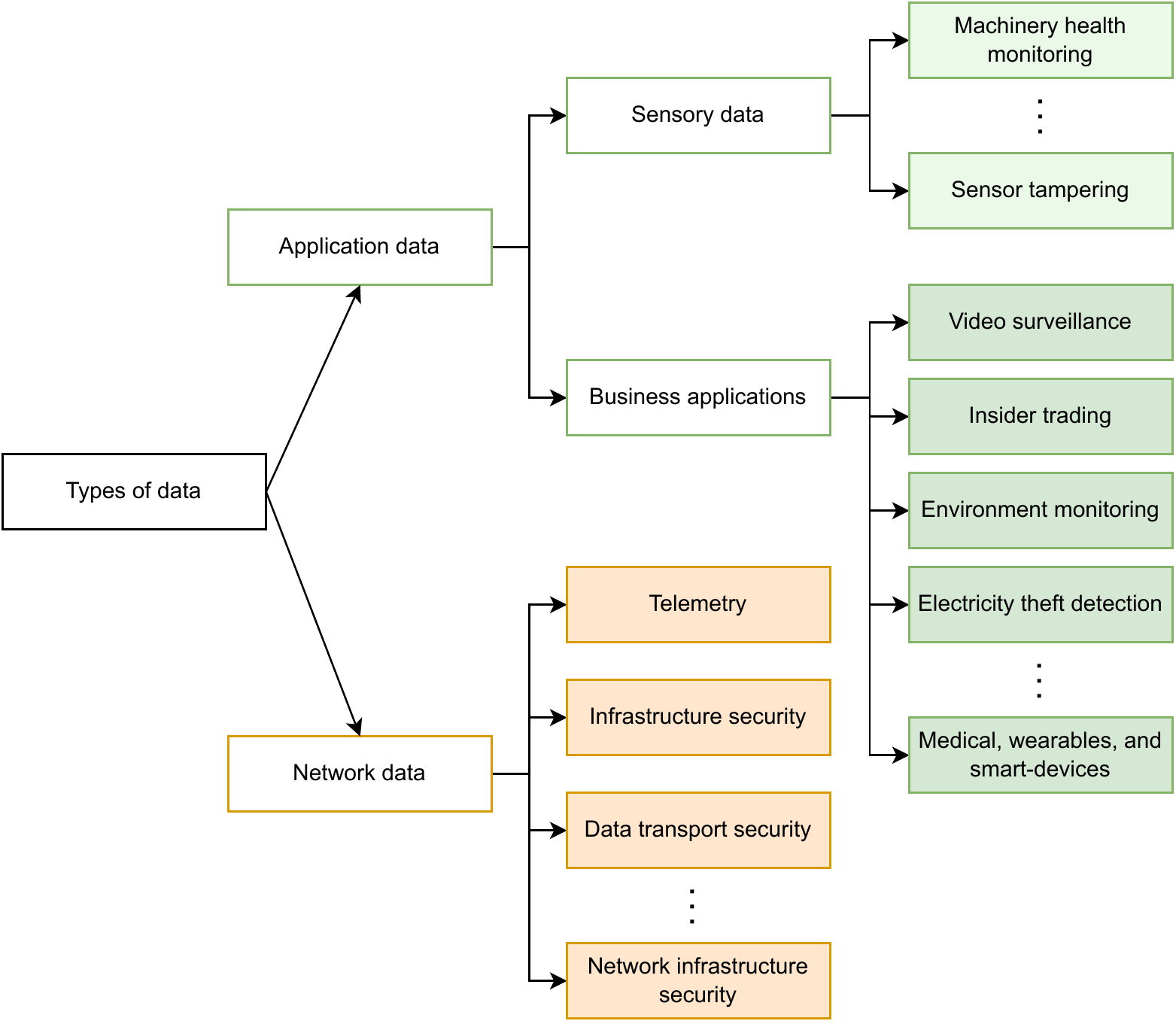}
    \caption{The two types of data (network and application data) and their applications.}
    \label{fig:dataTypes}
\end{figure}

This section covers the different applications of IoT anomaly detection from the selected papers. Application data and network data are the two main types of data. The data can be streamed in the case of time-series or processed from tabular data. Application data refers to data from IoT sensors processed to serve practical and business applications. Additionally, network telemetry data, such as CPU and memory usage, are monitored for the health of the IoT network. We found applications that are for individual or residential use and industrial applications. There are health monitoring applications for individuals that use data from wearable devices and smart home applications for residential use-cases. A few examples of industrial applications include monitoring the health of manufacturing equipment, ensuring the quality of data transferred through sensors, and smart city applications such as traffic monitoring. All of these applications require data security, user privacy, and reliability of data transfer. Figure \ref{fig:dataTypes} shows the two types of data and their respective applications.

On the basis of these data, we combined the different applications into sixteen categories. The categories and the papers are as follows.

\begin{table}[!ht]
\centering
\begin{tabular}{|r|l|} \hline
\textbf{Application Category} & \textbf{Reference} \\ \hline
Time-series data & \cite{HCLSTM2020, ElLoad2019, CBAD2020, WAAD2020, HADIoT2020, CNNLSTM2020, CLSTM2020, Stray2020}\\
Smart city & \cite{PPG2021, ILSTM2020, FoodWaste2020, SMS2020}\\
Monitoring machinery health & \cite{MT2020, machineMonitor2020}\\
Robotics and manufacturing & \cite{VAE2020_sliding, WFCM2020, GOil2020}\\
Detection applications for IoT sensors & \cite{IoTCAD2020, ST2019}\\
Other industrial or manufacturing & \cite{SAP2020, ConvLSTM2021, LogEvent2vec2020, MSC2021}\\
Surveillance and video & \cite{ImagePerSmoke2019, READIoT2021, CNNbiLSTM2021}\\
General-purpose frameworks & \cite{RAD2019, RELAD2019, WeatherView2020}\\
Network and communication frameworks & \cite{AnoMLIoT2021, DelIoT2021, Collaborative2020, EIHTTP2021}\\
User security and privacy frameworks & \cite{TSIndustrialIoT2020, MIoT2021, SDN2021}\\
Other frameworks & \cite{IWSN2020, XGBoostClassify2020}\\
Network traffic in IoT & \cite{TONTA2021, GEERDLAD2021, secureSVM2019}\\
Device and infrastructure security & \cite{SensorTampering2021, DIoT2019, DIoT2021, GraphDist2021, PbTA2021, ADSHospital2021}\\
Data transport security & \cite{HybridCyber2021, HMM2019, Botnet2020, DPI2020}\\
Other security applications & \cite{VAE2020, DAD2021, IDCSCNN2020, Fractal2020, APAE2021, CBRM2019, SEM2021, MFEWBagging2021, HiDim2020}\\
Miscellaneous IoT applications & \cite{ITrust2020, UgMining2021, horseBetting2021, HSTCN2021}\\ \hline
\end{tabular}
\caption{A table listing the 64 papers included in this review, as well as their respective application categories.}
\end{table}

\subsection{IoT anomaly detection applications with time-series data}

The first category is concerned with general time-series or streaming data, which are used in a range of applications. Hierarchical clustering, a non-parametric and lightweight approach, identifies linked sensors and generates clusters. It is possible to utilize a single detection method for all sensors in a cluster. For anomaly detection, the article by Shukla et al. \cite{HCLSTM2020} used robust statistical M-estimators paired with the LSTM neural network. Temperature data collected by local weather stations frequently contain missing values and inaccurate readings. The paper by Sobhani et al. \cite{ElLoad2019} shows a substantial relationship between temperature and load. The accuracy of final load projections is improved in the research by eliminating discovered observations from the original input data using local load information. The paper by Ngo et al. \cite{CBAD2020} demonstrated that using the proposed method in distributed hierarchical edge computing (HEC) systems reduces the detection delay by approximately 71\% while the accuracy is maintained. Anomaly detection models for univariate data in the model are built by using AEs. To associate with the three corresponding levels of HEC, the researchers proposed three AE-based models—AE-IoT, AE-Edge, and AECloud. The models have three, ﬁve, and seven layers to learn different features for data representation.

The wavelet autoencoder anomaly detection (WAAD) technique \cite{WAAD2020} initially uses a discrete wavelet transform to produce wavelet transform coefficients by applying a sliding time window to a time-series. The method is applicable on non-stationary and non-periodic univariate time-series data. It then encodes and decodes (reconstructs) these coefficients using an AE. For every time window, WAAD determines the reconstruction error and obtains to detect anomalous data. To identify local and global anomalies, Chang et al. designed a hierarchical anomaly detection architecture in \cite{HADIoT2020}. High detection accuracy is obtained by combining both edge and cloud servers, respectively. Local anomaly detection focuses on the consistency of the data pattern of individual devices using the Gated Recurrent Unit, which is subsequently sent to the cloud where the global anomaly detection procedure is conducted. The Global IoT anomaly detection looks at how various IoT devices communicate data and uses Conditional Random Fields for detection.

The paper by Yin et al. \cite{CNNLSTM2020} proposes a new architecture that integrates the convolutional neural network (CNN) and LSTM-based AE. The data from all sensors are collected and delivered to the processing center. Temporal features are extracted using a two-stage sliding window, and higher-level spatial features can be extracted using CNN. The LSTM is then applied to extract high-level temporal characteristics. On the other hand, the C-LSTM model \cite{CLSTM2020} consists of CNN, LSTM-based AE, and deep neural network (DNN) for anomaly detection. This technique applies two-stage sliding window processing to the data for high-level feature extraction. The reconstruction error is used to estimate anomalies in the method.

The HDoutliers algorithm is a powerful unsupervised method for discovering abnormalities in high-dimensional data based on a distributional model that allows outliers to be tagged with a probability. The algorithm tackles the curse of dimensionality, scalability, categorical variables, and non-normal distributions. To overcome the shortcomings of HDoutliers, the paper by Talagala et al. \cite{Stray2020} offers an improved technique for anomaly detection that uses higher-dimensional data, called the stray algorithm.


\subsection{Monitoring the health of machinery}
In contrast to the previous category, which looked at general time-series data, this category monitors the application data flowing from IoT sensors attached to industrial machinery and seeks machine faults. The proposed Mahalanobis–Taguchi method \cite{MT2020} is used to detect changes in the mechanical condition of the facilities. This technique can be used to keep an eye on the level of fatigue in logistic systems. Researchers in the paper have studied ways to detect damage to rotating equipment, such as bearings and gearboxes. By using the Mahalanobis distance, the method can calculate the feature vectors of the anomaly vibration data. A condition-based maintenance (CBM) approach can minimize scheduled and unscheduled breakdowns, as well as losses in production. CBM's primary purpose is to spot anomalies and anomalies that don't conform to any explicit laws and monitor those anomalies while they are still in progress. The Narrow Band Internet of Things (NB-IoT) technique is used in Li et al.'s work \cite{machineMonitor2020} to design a condition monitoring system for machinery equipment. In this approach, sensor data is sent to a server that then locates the cause of the anomalies in time and decides how much damage has been done. It uses wavelet packet decomposition to extract the features of the vibration signal.

\subsection{Robotics or manufacturing applications}
The third category inputs application data for anomaly detection in robotics and manufacturing. In the paper by Chen et al. \cite{VAE2020_sliding} a real-time anomaly detection system for industrial robots was created using a variational AE (VAE). The method was tested by injecting faults into the robot and observing how the robot resolved them. To cope with multiple time-series data, the proposed model can detect anomalies spatially and temporally. Another novel method to detect anomalies in the manufacturing process in \cite{WFCM2020} implements an unsupervised feature reduction method based on mutual information (MI) and conditional mutual information (CMI). A clustering model based on density peak (DP)-weighted fuzzy C-means (WFCM) is proposed to detect anomalies. An anomaly detection method for gearbox oil temperature utilizing SCADA data is proposed in \cite{GOil2020}. In the process, domain knowledge such as wind speed, power, low/high-speed shaft temperature, and intake oil temperature, respectively, Pearson correlation and Spearman rank correlation are used to measure the association between gearbox oil temperature and other directly related factors or indirectly related.

\subsection{Detection applications for IoT sensors}
The IoT-CAD method \cite{IoTCAD2020} proposes a context-aware adaptive data-driven monitoring system for IoT systems that can detect changes in sensor signals caused by an unexpected event in the environment. The process considers the correlation between the levels of oxygen, ammonia, and nitrate in the water tanks. The model is made to be adaptable and able to self-update. By monitoring changes in temperature and pressure, the model can detect anomalous rises in nitrate levels. Alternatively, Cui et al.'s approach \cite{ST2019} uses Spatio-temporal correlation to detect anomalies in IoT sensors. The method has two stages: (i) the anomaly source identification stage is completed by a fuzzy logic system based on Spatio-temporal correlation, and (ii) the anomaly detection stage using a composite distance metric and sensor clustering. The approach exploits the temporal sequence to get the historical spatial correlation degree, i.e., historical feature parameters.

\subsection{Other industrial or manufacturing applications}
Other approaches include merging edge devices' electricity data with weather data allows \cite{SAP2020} method to create a sparse anomaly perception model. By following this method, anomalous data can be labeled quickly and efficiently. Navigation and planning tasks are particularly vulnerable to driving errors in autonomous vehicles. The safety and security of connected autonomous vehicles' passengers are essential for the development of autonomous vehicles. The method by Zekry et al. \cite{ConvLSTM2021} proposes an IoT sensor-assisted convolutional long short-term memory (LSTM) model for connected vehicles for anomaly detection. The method by Wang et al. \cite{LogEvent2vec2020} targets log anomalies in large-scale IoT systems. Using a natural language processing approach, the researchers extracted the relevance between words and vectorized them. The method trains supervised models (Random Forests, Naive Bayes, and Neural Networks) to detect anomalies and cites a 30x reduction in computational time. Dang et al. \cite{MSC2021} proposes a novel monotone split and Conquer (MSC) strategy to detect short and long forms of anomalies. The MSC model has an offline training phase and an online detection phase. The method extracts sub-trends from the data and uses the multi-scale PCA algorithm to generate a normal data profile.

\subsection{Smart-city IoT applications}
Smart-city and urban applications include the paper by Xu et al. \cite{ILSTM2020} proposed an integrated IoT anomaly detection method, dubbed I-LSTM, presented based on the concept drift adaptive and deep learning methods to detect anomalies in smart-city data. The effectiveness of I-LSTM is with a smooth activation function for multi-classification anomaly detection, which can obtain the specific anomaly classification to improve the integrated quality of service of a smart home. To differentiate valid and noisy Photoplethysmography (PPG) signals collected from wearable devices, the quality assessment lightweight PPG method \cite{PPG2021} employs the Mahalanobis distance measure. The approach is non-invasive and is suitable for monitoring vital indicators such as heart rate, breathing rate, and blood oxygen saturation without subjecting the patient to procedures that could become unnecessary.

Smart cameras and temperature-meters tied to an edge gateway at quick service restaurants predict demand, creating timely alerts and intelligent decisions for proactive waste management. The approach by Aytaç and Korçak \cite{FoodWaste2020} is one such application to detect sensor defects, monitor security attacks, and detect increased foot traffic in restaurants in the edge environment of IoT. The stored data are first broken down into separate groups with the K-means clustering or Lloyd's method. Following the implementation of the clustering method, the experiment is executed to determine the correctness of the clustering results. Then anomalies are discovered based on the Bhattacharyya distances. Another application is in the medical sector to protect the safety and privacy of the patient's personal and medical data in a smart city, the paper by Tripathi et al. \cite{SMS2020} proposed the Smart Medical System (SMS) method. Classical healthcare systems were shown to have notable shortcomings and their incompatibility with smart city systems. The proposed framework suggests safe, rapid, reliable, and transparent city-wide connectivity based on IoT, MEC, and blockchain consensus processes. Because the data need to travel only a small distance between the sensor and the edge nodes, it is faster, uses less energy, and is more cost-effective with fewer data losses.

\subsection{Surveillance and video IoT applications}
This category includes IoT anomaly detection algorithms that make use of application data from cameras and remote sensing sensors. Inadequate monitoring of electrical equipment can result in massive economic losses and societal impacts. Hou et al. \cite{ImagePerSmoke2019} proposed an image anomaly detection method for IoT equipment based on deep learning for both personnel identification and fire smoke detection to address the problem. The researchers used the approach to monitor the safety of the power equipment's working environment and people identification, and fire smoke detection. The main contribution includes introducing color features, texture features, and shape irregularities to increase the recognition rate. Furthermore, based on deep convolutional neural networks, the algorithm detects fire smoke.

Another is a surveillance system developed for hostile environments for the Tunisian army, dubbed Read-IoT \cite{READIoT2021}. The project's primary goal is to design a system for monitoring security threats that is both dependable and complete. The proposed system, READ-IoT, applies to IoT networks of heterogeneous objects such as cameras, sensors, and drones. The READ-IoT framework responds to real-time constraints and helps to reduce system downtime for cloud or fog computing. Threat detection is handled automatically and appropriate responses are executed. The approach requires cascaded activation of detection components. Data from each subsystem (anomaly detection system and event detection system) are funneled into a centralized source for effective decisions in the field.

The third is a proposed framework by Ullah et al. \cite{CNNbiLSTM2021} that reduces the amount of time required to identify abnormal occurrences in surveillance networks. The proposed system incorporates bi-directional LSTM with CNN characteristics to identify and classify anomalous incidents in the real world. In this pipeline, CNN features are collected from successive frames, and then a new multi-layer LSTM is used to distinguish the normal and abnormal. The deep features and multi-layer BD-LSTM provide flexible, high-level training and validation data to real-world surveillance networks. This framework might be used by law enforcement at airports and hospitals, for example.

\subsection{General-purpose frameworks}
With network and application data, the following four categories outline the recent frameworks and their applications. General-purpose frameworks include the robust anomaly detection framework (RAD Framework) \cite{RAD2019} developed for less reliable data. The framework utilizes SVM, KNN, random forest, and nearest centroid classifiers. The authors claim to achieve an accuracy of up to 98\% for attacks on IoT devices and cluster failure prediction, and comment that the accuracy reduction is due to general reduction in clean data and noise pollution. For industrial IoT applications, Wang and Ahn \cite{RELAD2019} found that monitoring current power consumption is not sufficient to meet the second-by-second balance between power consumption and generation. To approach the problem, the article proposes a hybrid one-step-ahead load predictor (OSA-LP) and a rule-engine-based load anomaly detector (RE-AD), a framework that could detect anomalies in residential power usage in real-time. Koduru et al. \cite{WeatherView2020} proposed another framework that creates a pipeline for a weather monitoring system. The framework uses a few sensors to provide information on any unexpected weather changes, and the findings are displayed on a smartphone.

\subsection{Frameworks for network applications}
The frameworks that make use of network data fall into the following group. From a network, network security, and communication standpoint, AnoML-IoT \cite{AnoMLIoT2021} is a data science pipeline that integrates several wireless communication protocols, anomaly detection algorithms, and deployment to edge, fog, and cloud platforms with little user input. The pipeline covers four major phases: data intake, model training, model deployment, inference, and maintenance. The DeL-IoT framework \cite{DelIoT2021} for IoT anomaly identification and prediction uses a deep ensemble learning technique to identify anomalies. The framework detects IoT abnormalities by dynamically observing packet and flow level traffic instances that pass through SDN switches and system metrics. In the scalable framework by Mirsky et al. \cite{Collaborative2020}, a distributed and collaborative anomaly detection algorithm is created using the concept of blockchain. The method uses a probabilistic model called a Markov chain (MC) to simulate sequences efficiently. An et al. \cite{EIHTTP2021} proposed a unique anomaly detection framework for the IoT, which is capable of relieving network congestion and central processing units (CPUs) from the computing pressures of centralized servers while unlocking the potential of edge intelligence (EI) in the IoT. The framework uses clustering and classification algorithms sequentially.

\subsection{User security and privacy preserving frameworks}
Frameworks that incorporate user security and privacy, like the paper by Liu et al. \cite{TSIndustrialIoT2020} proposes a framework for on-device personal data processing and handles user data locally. An attention mechanism-based neural network is used in this on-device federated learning framework. As part of this system, each edge device utilizes the local dataset to train the global model supplied by the cloud aggregator and then transfers the gradients back to the cloud aggregator until convergence. Decentralized Industrial IoT devices might make use of global anomaly detection algorithms in this way. The MIoT framework \cite{MIoT2021} provides a multi-dimensional view of the anomalies and the accompanying issues. The model considers two kinds of content anomalies: tight content anomalies, in which the complete set of reference keywords must be included in the related transactions, and loose content anomalies, in which at least one of the reference keywords must be present. Another paper \cite{SDN2021} proposes a more efficient approach for detecting malicious behavior in SDN and edge computing networks. The method helps verify the trustworthiness of edge devices for data forwarding while maintaining data secrecy throughout data transmission and exchange.

\subsection{Other frameworks}
Other frameworks like the paper by Li et al. \cite{IWSN2020} use sensor processing, smart meter readings, and a blockchain to detect anomalies in electricity use accurately and on time. Using the system, factories and residents can be warned to reduce electricity use. Anomaly detection in IoT data is performed using a service selection framework by Yang et al. \cite{XGBoostClassify2020}. The model utilizes a rapid classification algorithm, which is an XGBoost implementation that is taught to recognize distinct stream data patterns and allow for the training of new decision models.

\subsection{Network traffic in IoT}

Network traffic analysis (NTA) is a specialized topic within network traffic prediction. In ad hoc IoT networks, such as when using ML-based NTM methods, training is time-consuming, and unexpected occurrences or departures are often missed. TONTA \cite{TONTA2021}, a suggested online anomaly and trend change detection technique, is used by all nodes responsible for data forwarding. Although TONTA identifies the predominant trend changes, some events can cause jitters, e.g., performance of communication protocols and network congestion, to name a few. The Green Energy Efficient Routing with Deep Learning Based Anomaly Detection (GEERDLAD) model \cite{GEERDLAD2021} provides an effective use of energy to help increase the network span. Anomaly detection in the IoT communication networks is carried out in the algorithm through the recurrent neural network-long short term memory (RNN-LSTM) model. Furthermore, the paper by Shen et al. \cite{secureSVM2019} proposes a privacy-preserving SVM training strategy, called secure SVM. The method uses encrypted IoT data, in which data providers encrypt their data locally using their private keys and then record the encrypted data on the blockchain through specially structured transactions.

\subsection{Security applications for IoT devices and infrastructure}

The IoT anomaly detection community is focused primarily on security applications. Because of the large number of publications in this sector, it is not possible to cover all of them; nonetheless, we have categorized security applications into three categories - (i) devices and infrastructure, (ii) data transport, and (iii) miscellaneous applications. This section discusses the first of the three categories, devices and infrastructure, which use a combination of streaming application data and network data to identify security anomalies affecting connected hardware devices.

In Pathak et al.'s paper \cite{SensorTampering2021}, the researchers used the isolation forest approach to detect sensor tampering. Sensor tampering is detected using network traffic data and is applicable in an industrial and office environment. The algorithm takes packet length, packet-length-daily count, and hash value as input, along with the Silhouette coefficient metric for evaluating the goodness of a clustering technique. A distributed and self-learning approach, DIOT is an autonomous, self-propelling system that aims to find compromised IoT devices. It automates device-specific communication profiles built on through device-type-specific communication profiles with no human intervention or labeled data, which are used to detect anomalies in devices' behavior. The paper by Nguyen et al. \cite{DIoT2019} has a dedicated model for each device type for different IoT devices with different behaviors. To improve global model accuracy, the DIOT approach by Yahyaoui et al. \cite{DIoT2021} has an ensembler part aggregating changes from multiple sources and then optimizes the model's accuracy. Instead of having two gates for each memory cell, they only use two gates in total, the Reset gate and Update gate. The approach is for AI-powered anomaly detection on IoT networks, utilizing federated learning. The system has a Virtual Instance, a local deep learning model, and an ensembler component with a high-level architecture.

The design of a distributed anomaly-detection system involves implementing a graph neural network (GNN) method \cite{GraphDist2021}. This system is designed to thoroughly monitor all of the network's infrastructure. GNN monitoring is performed to monitor nodes and devices connected to those nodes using the underlying graph structure. To protect private data from home devices, Venkatraman et al. \cite{PbTA2021} used a Probabilistic Timed Automaton (PTA) to model the activities of smart devices, keeping intruders from stealing confidential data. Furthermore, Said et al. \cite{ADSHospital2021} use the support vector machine (SVM) classifier to help preserve data accuracy while avoiding sending false alerts to hospital infrastructure. Body temperature and heartbeat data are among the applications.

\subsection{Security application for IoT data transport}

The second category of IoT anomaly detection for security applications is data transport. The methods used in this application provide secure and uncompromised transit of application data through the network and identify malleable attacks such as man-in-the-middle and trojans. Niraja et al. \cite{HybridCyber2021} propose an adaptive hybrid strategy for near-real-time detection of IoT cyberattacks to improve IoT security. It is based on the integration of deep AE and feature extraction. During the process, it performs encoding and decoding to have the data compressed and reconstructed, respectively, to detect anomalies or attack traffic generated by compromised IoT devices. Fouad et al. \cite{HMM2019} propose a technique to detect attacks, trojans, and malfunctions. It is beneficial for the detection of attacks with low training samples. Hidden Markov models (HMM) are used in all aspects of the model, and power signature analysis is used to detect anomalies. Developing new methodologies to discover compromised IoT devices is imperative as there will be significant impacts if IoT botnet attacks are left unmanaged. The proposed method by Shorman et al. \cite{Botnet2020} makes the main contribution in botnet attack detection using the Grey Wolf Optimization algorithm. The Song et al.'s technique \cite{DPI2020} uses a deep packet inspection (DPI) approach and is based on the Three Sigma Rule and Hurst parameter. Using DPI-based solutions, network operators can provide a detailed view of network usage, identify heavy users, and respond quickly to network traffic. This is useful for finding and preventing network abnormalities that lead to increased use of router resources.

\subsection{Other security applications}

The third security category is miscellaneous and covers applications that indirectly affect IoT sensors, infrastructure, application, network, or ML models. For example, in MLOps, where continuous delivery and execution of ML software in containers such as Docker containers running in Kubernetes are required, network data are utilized to evaluate whether an unexpected container shutdown is from a security attack. In a distributed denial of service (DDoS) attack, the network is bottlenecked by incoming handshaking requests.

The article by Vu et al. \cite{VAE2020} presents a learning approach that inherits the strength of supervised learning approaches to identify known threats to VAEs. VAEs learn to map incoming data into a single area in its bottleneck layer using the standard Gaussian shape. The experimental findings indicate that the proposed models can map nonlinear separable normal and attack data in their original space to linear and isolated data in their latent feature space. Wang's paper \cite{CBRM2019} suggests a network-monitoring algorithm built around a cluster-based routing method. Cyber-attacks like DDoS attacks are detected with this method. The credibility level is evaluated, and anomalous activity is monitored with a trusted function module in this approach.

The edge network, including routers and switches in locations such as the airport and gateway, is safeguarded with a proposed ensemble one-class statistical learning model by Moustafa et al. \cite{DAD2021} that implements Gaussian Mixture-based Correntropy. The proposed system aims to identify zero-day attacks and also attempts to develop a legitimate profile for new data flows as they occur. Data mining methodologies that include clustering and rule discovery, statical approaches, and machine learning approaches have been the core hierarchy for anomaly detection-based intrusion detection systems. The proposed system called IoT-based Intrusion Detection and Classification System using Convolutional Neural Network (IoT-IDCS-CNN) \cite{IDCSCNN2020} has three subsystems: a feature engineering subsystem, a feature learning subsystem, and a traffic classification subsystem. The paper by Dymora and Mazurek \cite{Fractal2020} used fractal analysis to identify potential security problems in a network. Based on the findings of the paper, it appears that it would be possible to detect both short-term attacks and more intense ones. Basati and Faghih's approach \cite{APAE2021} proposes a real-time network intrusion detection system that employs two AEs in parallel to monitor the network and a feature reduction deep AE to identify the most distinct features.

A stacked ensemble meta-learning (SEM) model \cite{SEM2021} is proposed to boost the effectiveness of the base machine learning model for IoT device anomaly detection. This design builds a higher-level prediction model that encompasses the predictions of base classifiers that have low accuracy. To have a high-level prediction, the approach utilizes a meta-learning prediction approach. The goal of Bhatia and Sangwan's paper \cite{MFEWBagging2021} is to make debugging and explaining easier using fewer features. The approach uses an ensemble feature selection to eliminate the bias of individual feature selection methods during the ensemble and identifies the optimal subset with non-redundant and relevant features.

Furthermore, Kurt et al. \cite{HiDim2020} propose extracting useful univariate summary statistics from observed high-dimensional data and performing anomaly detection in a single-dimensional space. Suppose that the observed data have low intrinsic dimensionality. In that case, the method learns a submanifold in which the nominal data are embedded and determines whether the sequentially acquired data persistently deviate from the nominal submanifold. In the general case, Geometric Entropy Minimization is used to learn an acceptance region for nominal data and evaluate whether the sequentially observed data consistently fall outside the acceptance region.

\subsection{Miscellaneous IoT applications}

IoT anomaly detection is still a developing field, with some applications receiving fewer publications than others. The last category, miscellaneous, includes applications in their early stages of development. This includes the paper by Liu et al. \cite{UgMining2021}, which discussed the data acquired by several mining sensors and the requirement for real-time tracking of safety warnings. Construction safety monitoring is performed to look for inconsistencies in the data gathered by the sensors. In this approach, anomaly detection tasks are separated from the sensor and sink nodes and assigned to distinct nodes in the network. Then, algorithms are built for each kind of node, allowing for more processing flexibility. In noisy environments, the ITrust model \cite{ITrust2020} can detect anomalous or faulty nodes more effectively than current trust models. This model is primarily intended for underwater acoustic sensor networks. The ITrust, which contains four trust metrics (communication trust, data trust, energy trust, and environmental trust), is built on the isolation forest algorithm. The paper by Min et al. \cite{horseBetting2021} proposes a framework for Mobile Horse Racing Betting (MHRB) relying on IoT sensors, devices, and applications. The framework considers recurrent neural network (RNN), long short-term memory (LSTM), and statistical methods to discover insider trading. Furthermore, as IoT devices collect an enormous amount of communication records, the data rapidly. In Cheng et al.'s paper \cite{HSTCN2021}, a semi-supervised model was developed to identify anomalies in such scenarios. The method uses a stacking approach, which is an effective way to evaluate data without a label. It is used to discover records with uncertain data and eliminate them.

The section concludes with a table in \ref{tab:mostcitedpapersrecent} of fifteen articles having a distinct application domain and with the highest citation count from 2019 until July 2021 (five for each year).

\begin{table}[!ht]
\centering
    \begin{subtable}[h]{\textwidth}
        \centering
        \begin{tabular}{|c|c|c|c|} \hline
        Paper & Application & Publisher & Cites \\ \hline
        \cite{Botnet2019} & Botnet attack detection & Elsevier & 253 \\
        \cite{Blockchain2019} & Blockchain-based encrypted IoT data & IEEE & 164\\
        \cite{DIoT2019} & Federated approach for smart home & IEEE & 131\\
        \cite{SDN2019} & Flow detection in software-defined networks & IEEE & 112\\
        \cite{Cyberman2019} & Attacks in cyber manufacturing systems & Springer & 110\\ \hline
       \end{tabular}
       \caption{2019}
       \label{tab:mostcited2019}
    \end{subtable}
    \hfill
    \begin{subtable}[h]{\textwidth}
        \centering
        \begin{tabular}{|c|c|c|c|} \hline
        Paper & Application & Publisher & Cites \\ \hline
        \cite{GARUDA2020} & Approach to intrusion detection & Springer & 73\\
        \cite{Botnet2020} & IoT botnet detection & Springer & 56\\
        \cite{VLSTM2020} & Industrial big data applications & IEEE & 52\\
        \cite{Security2020} & Gray hole attack detection & Springer & 48\\
        \cite{Trustworthy2021} & Framework to determine trustworthiness & IEEE & 44 \\ \hline
        \end{tabular}
        \caption{2020}
        \label{tab:mostcited2020}
     \end{subtable}
     \begin{subtable}[h]{\textwidth}
        \centering
        \begin{tabular}{|c|c|c|c|} \hline
        Paper & Application & Publisher & Cites \\ \hline
        \cite{Cvitic2021} & IoT generated DDOS traffic detection & Springer & 29\\
        \cite{CNNbiLSTM2021} & Detection in surveillance networks & Springer & 22\\
        \cite{IDBlockchain2021} & Blockchain-based systems intrusion detection & IEEE & 22\\
        \cite{IoT5G2021} & Cybersecure IoT and 5G infrastructure & Elsevier & 16\\
        \cite{SmartCity2021} & Multiple IoT scenarios in a smart city & Elsevier & 15 \\ \hline
        \end{tabular}
        \caption{2021}
        \label{tab:mostcited2021}
     \end{subtable}
     \caption{Fifteen papers with the highest citation count (as of 5th of August 2021) are listed on this table (five for each year). They are grouped by year and has a unique application domain.}
     \label{tab:mostcitedpapersrecent}
\end{table}

\newpage
\section{Conclusions and new perspectives} \label{sec:conclusion}
The frequency of publication in IoT anomaly detection shows that the field is still in its early stages. Anomaly detection algorithms are classified into four categories, which are briefly summarized in this review. It also lists the most commonly used keywords and applications and identifies application domains that require further research. Lastly, based on our search criteria, this review discusses 64 novel papers published between January 2019 and July 2021, each with a distinct application domain. As per our current knowledge, there is no single best generic algorithm for the problem but rather a number of methods specific to a particular application. In this section, we attempt to narrow down a large number of possibilities to some of the most significant issues facing the field today. The new perspectives are beneficial for univariate, multivariate, as well as for dealing with high-dimensional data. They are:
\begin{itemize}
    \item \textit{Unsupervised, semi-supervised, and self-supervised approach:}\\
    Obtaining labeled training data is time-consuming, expensive, and not always possible. Furthermore, the labeled data may not accurately represent all anomalous data present in the dataset. There may also be an imbalance between the types of anomalous data present for training and testing. As a result, a self-supervised, semi-supervised, and unsupervised algorithm, especially as an online approach, that can achieve detection accuracy comparable to that of supervised approaches is desirable.
    \item \textit{Detection for multi-representation data traffic:}\\
    The representation data from one or more types of sensors may be sent over the same data pipeline. One such use is external building profiling in an urban environment \cite{multirep2021}, which involves the integration of RGB cameras with sensors such as hyperspectral, LiDAR, and thermal, among others. A detection algorithm capable of handling volumes of multi-representation data flow at scale and with great precision and recall is viable for these types of applications.
    \item \textit{Drift adaptation:}\\
    A factor in the accuracy of a trained model is the distribution and quality of the data available during the training process. Over time, when the distribution of the incoming traffic differs from the distribution of the training data, there is a data drift \cite{DriftReview2019}. In such cases, it is common for the model to require re-training. It is preferable for an anomaly detection model to take a more proactive strategy to adapt to incoming data or concept drifts.
    \item \textit{Learning and testing from augmented data:}\\
    In order to train a classification algorithm, it is necessary to have enough accurate and reliable data. This is especially true for the binary classification of anomaly detection. However, when there are insufficient data available for one or more classes, statistical and machine learning-based data augmentation methods such as the generative adversarial network (GAN) generate more training samples \cite{GAN2021}. A robust detection technique that utilizes data augmentation in model training and software testing is required for applications with insufficient data.
\end{itemize}

\section*{Acknowledgments}
This work has been funded by the Knowledge Foundation of Sweden (KKS) through the Synergy Project AIDA - A Holistic AI-driven Networking and Processing Framework for Industrial IoT (Rek:20200067).

\bibliographystyle{unsrt}  
\bibliography{manuscript.bib}

\end{document}